\documentclass{article}

\usepackage{arxiv}

\usepackage[utf8]{inputenc} 
\usepackage[T1]{fontenc}    
\usepackage{hyperref}       
\usepackage{url}            
\usepackage{booktabs}       
\usepackage{amsfonts}       
\usepackage{nicefrac}       
\usepackage{microtype}      
\usepackage{lipsum}		
\usepackage{graphicx}
\usepackage{natbib}
\usepackage{doi}
\usepackage{amsmath}

\usepackage{fancyhdr}
\usepackage{textcomp}

\fancypagestyle{plain}{
  \fancyhf{}
  \fancyfoot[C]{%
    \begin{minipage}{1\textwidth}
    \raggedright
    \footnotesize
    \textcopyright\ 2026 Annie Yihong Yuan. All rights reserved. \\
    Expert Cognition Dashboard\texttrademark\ is a trademark claimed by the author. 
    All figures and dashboard mock-ups are original works.
    \end{minipage}
  }

}

\title{Expert Cognition Dashboard:
From Learning Analytics to Cognition Intelligence in AI-Driven Education}

\author{ 
	\href{https://orcid.org/0009-0004-1760-0149}{\includegraphics[scale=0.06]{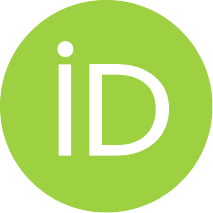}\hspace{1mm}Annie Yuan} \\
	School of Computer Science\\
	The University of Sydney\\
	NSW, 2006, Australia \\
	\texttt{annie.yuan@sydney.edu.au} \\
}

\renewcommand{\shorttitle}{Expert Cognition Dashboard}

\hypersetup{
pdftitle={A template for the arxiv style},
pdfsubject={q-bio.NC, q-bio.QM},
pdfauthor={David S.~Hippocampus, Elias D.~Striatum},
pdfkeywords={First keyword, Second keyword, More},
}

\begin{document}
\maketitle
\thispagestyle{plain}

\begin{abstract}
Current AI-driven educational systems primarily rely on behavioural analytics, performance metrics, and content-level interactions to model learning processes. While these approaches provide measurable indicators of learner activity, they remain insufficient for representing the deeper cognitive structures that experts use when interpreting learner development, identifying misconceptions, and making adaptive pedagogical decisions. Existing learning analytics dashboards are largely designed to visualise learner behaviour for human instructors, rather than to embody expert cognition as a reasoning infrastructure for AI-native education.
This paper introduces the \textit{Expert Cognition Dashboard\texttrademark} (ECD), a cognition-centred reporting infrastructure for AI Twin-driven education systems. The central aim of ECD is to model expert cognition and operationalise it within dashboard systems, enabling learner behaviours to be interpreted through expert-like cognitive structures rather than treated as raw behavioural signals. Instead of allowing AI systems to directly reason from learner interaction traces, the proposed framework transforms student interactions into interpretable cognition structures through AI Tutor analysis and multi-level dashboard aggregation. The architecture organises cognition across three interconnected layers: individual cognition dashboards, class cognition dashboards, and AI Twin expert dashboards for cross-group reasoning and adaptive intervention.
Building upon the framework of AI Expert Feedback Ecology, the proposed system redefines dashboards as cognitive middleware that connects learner behaviours with AI-driven expert reasoning. By modelling dimensions such as interpretation, identity cognition, value recognition, misconception patterns, and learning tension, ECD enables AI Twins to identify recurring learner difficulties, generate adaptive tasks, and support personalised intervention strategies.
This paper argues for a paradigm shift from learning analytics toward Cognition intelligence, positioning dashboards not merely as visualisation tools, but as foundational cognition infrastructures that embed expert reasoning into future AI-native education systems.
\end{abstract}


\section{Introduction}

Recent advances in large language models (LLMs), generative AI, and adaptive learning technologies are rapidly transforming the educational landscape \cite{zhai2021review,kasneci2023chatgpt}. AI-driven tutoring systems are increasingly capable of providing personalised explanations, conversational support, automated assessment, and real-time feedback at scale. Emerging educational ecosystems now integrate AI Tutors, intelligent recommendation systems, multimodal learning environments, and agent-based interactions into continuous learning processes. As a result, education is gradually shifting from static content delivery toward AI-native learning environments, in which AI agents, learner models, adaptive systems, and data infrastructures are embedded into ongoing educational processes.

Within this transformation, AI Tutors and intelligent tutoring systems have long been central to AI in education, with systems designed to model learner knowledge, provide feedback, and adapt instruction to learner performance \cite{anderson1995cognitive,vanlehn2011relative,aleven2010rule}. More recently, LLM-based systems have expanded the conversational and generative capabilities of AI tutoring, enabling systems to simulate dialogue, scaffold learning tasks, and respond to student questions with increasing fluency and contextual awareness \cite{stamper2024enhancing,kasneci2023chatgpt}. At the same time, adaptive learning systems aim to personalise educational pathways through behavioural tracking, recommendation algorithms, learner modelling, and performance prediction.

Despite these advancements, many existing AI educational systems remain fundamentally behaviour-centric. Current approaches often rely on measurable learning activities such as task completion, clickstream patterns, engagement duration, quiz performance, and interaction frequency. These approaches are closely aligned with traditions in learning analytics and educational data mining, which have developed powerful methods for collecting, modelling, and analysing learner data at scale \cite{siemens2012learning,baker2009state,ferguson2012learning}. Although such behavioural signals are useful for evaluating participation and short-term performance, they provide limited insight into the expert cognition required to interpret learner development, diagnose misconceptions, and make adaptive pedagogical decisions. Consequently, AI systems may respond to learning behaviours without adequately representing the expert-like cognitive structures needed to reason about the meaning of those behaviours.

Learning analytics dashboards have become a standard component of modern educational platforms, providing visual interfaces for monitoring learner activity, progress, and performance \cite{duval2011attention,verbert2014learning}. These dashboards commonly visualise indicators such as grades, completion rates, attendance, engagement scores, and interaction histories in order to support instructors and administrators in monitoring learning progress. Existing systems are effective at presenting quantitative summaries of learner activity and identifying surface-level patterns across individuals and groups. However, traditional learning analytics approaches also exhibit limitations. While learning analytics has been explicitly framed as being about learning rather than data alone, many practical implementations still privilege behavioural traces and measurable indicators over deeper accounts of cognition and expert interpretation \cite{gavsevic2015let}.

First, existing dashboards primarily focus on observable behaviours rather than interpretive cognition. Metrics such as scores and participation rates cannot adequately represent how learners conceptualise knowledge, negotiate uncertainty, construct meaning, or develop value-oriented judgments. Second, current dashboards are largely designed for human interpretation and institutional monitoring, rather than for AI-driven expert reasoning and adaptive cognition modelling. Third, behavioural visualisations often reduce learning into measurable outputs while overlooking the internal tensions, misconceptions, and interpretive dynamics that experts typically consider when evaluating learner development.

As AI systems become more deeply integrated into educational processes, the limitations of behaviour-centred analytics become increasingly significant. An educational AI system may detect that a student answered incorrectly, but it often cannot determine whether the error emerged from conceptual misunderstanding, interpretive conflict, value tension, identity uncertainty, or incomplete cognitive transfer. In this sense, current dashboards provide data visibility without cognition visibility, and learner analytics without an explicit model of expert interpretation.

Human experts do not evaluate learners solely through behavioural statistics. Expertise is often grounded in tacit inference, pattern recognition, situated judgment, and reflective action \cite{polanyi1966logic,dreyfus1986mind,klein2017sources,schon2017reflective}. Experienced educators, mentors, and domain specialists interpret learner development through deeper cognitive structures, including conceptual understanding, interpretive reasoning, value alignment, identity formation, and tension recognition. Such judgment is shaped by the activity, context, and culture in which knowledge is used \cite{brown1989situated,lave1991situated}. The value of expert cognition therefore lies not only in domain knowledge, but in the ways experts notice patterns, interpret information, identify meaningful tensions, and decide what forms of support are pedagogically appropriate.

This reveals a critical missing layer in current AI educational systems: expert cognition-level interpretation. Although AI Tutors and learning analytics platforms can collect large amounts of behavioural data, they generally lack mechanisms for transforming those behaviours into expert-readable and AI-actionable cognition structures. As a result, AI systems remain limited in their ability to perform expert-like reasoning, adaptive intervention, and meaningful pedagogical reflection.

To address this gap, AI education systems require a new form of infrastructure capable of translating learner behaviours into interpretable cognition representations grounded in expert reasoning. Rather than treating dashboards as passive visualisation interfaces, future systems must support cognition-aware interpretation that enables AI agents to reason about learners in ways that approximate expert educational practice.

This paper argues that AI Twins should not directly observe learner behaviours as raw data. Instead, they should reason through cognitively interpreted dashboards generated from AI Tutor analysis and structured by expert cognition models. Learner interactions should first undergo cognition-level interpretation through AI Tutor analysis and multi-layer aggregation mechanisms. The resulting cognition structures can then be organised into expert-readable dashboards that support adaptive reasoning, intervention, and educational decision-making. Under this perspective, dashboards are no longer merely visualisation tools. They become cognition-centred infrastructures that mediate between learner behaviours and AI-driven expert reasoning.

This paper makes five contributions. 
First, it introduces the concept of the Expert Cognition Dashboard (ECD) as a cognition-centred educational reporting infrastructure for AI-native learning systems. 
Second, it proposes the concept of Cognition intelligence, shifting educational analytics from behavioural measurement toward interpretable cognition structures grounded in expert reasoning.
Third, it defines dashboards as a form of Cognitive Middleware that connects learner behaviours with AI-driven expert reasoning and adaptive intervention. 
Fourth, it presents a three-level cognition architecture consisting of individual cognition dashboards, class cognition dashboards, and AI Twin expert dashboards. 
Fifth, it proposes the framework of AI Expert Feedback Ecology, describing a closed-loop educational ecosystem in which AI Twins, AI tutors, dashboards, and learners continuously interact through cognition-aware feedback processes.

\section{Related Work}

This section situates the Expert Cognition Dashboard (ECD) within four bodies of work: learning analytics dashboards, AI tutoring and adaptive learning, human--AI reasoning, and expert cognition. These areas provide important foundations for learner data visibility, adaptive educational support, interpretable AI systems, and situated educational judgment. However, they also reveal a shared gap: current systems can collect, visualise, predict, and respond to learner data, but they do not yet provide cognition-centred infrastructures that transform behavioural traces into interpretable structures for AI-driven expert reasoning.

\subsection{Learning Analytics Dashboards and Behavioural Visibility}

Learning analytics has emerged as a major field concerned with the collection, analysis, and interpretation of learner data to understand and support learning processes \cite{ferguson2012learning,siemens2012learning}. Within this field, learning analytics dashboards have become one of the foundational infrastructures in contemporary digital education environments. Existing dashboards commonly provide visual representations of learner activities, including performance scores, completion rates, attendance, engagement frequency, time-on-task metrics, and behavioural interaction histories. They are widely used in online learning platforms, intelligent tutoring systems, and institutional educational management systems to support monitoring, assessment, and decision-making processes.

Prior research has also examined the relationship between learning analytics and educational data mining, particularly in relation to predictive modelling, learner monitoring, and data-informed educational intervention \cite{siemens2012learning,baker2009state}. In dashboard design, visualisation and recommendation have been central concerns, with systems often designed to help educators identify at-risk students, evaluate engagement patterns, and improve instructional efficiency through quantitative indicators \cite{duval2011attention,verbert2014learning}. Visualisation techniques such as heatmaps, progress bars, participation timelines, and performance distributions have therefore become standard methods for representing learner activity and group trends.

Despite their practical utility, current learning analytics dashboards remain largely behaviour-oriented. Although learning analytics has been explicitly framed as being about learning rather than data alone \cite{gavsevic2015let}, many dashboard systems continue to treat learner behaviour as the primary source of educational interpretation. Measurable interactions are often assumed to sufficiently represent learning states, yet behavioural traces alone provide limited access to deeper cognitive processes such as conceptual interpretation, value negotiation, identity formation, and cognitive tension. Existing dashboards frequently emphasise what learners do, while providing limited insight into how learners think, interpret, or construct meaning.

This limitation becomes especially important when learning is understood as a social and interpretive process. Social learning analytics, for example, has highlighted the importance of interaction, discourse, and collective meaning-making in educational data practices \cite{shum2012social}. However, even socially oriented analytics systems often remain constrained by what can be captured, counted, and visualised as learner activity. As a result, dashboards may reveal patterns of participation or interaction without representing the cognition structures that give those patterns educational meaning.

In addition, traditional dashboards are predominantly designed for human instructors rather than AI-driven reasoning systems. Their primary objective is to support visual inspection and institutional reporting, not cognition-level interpretation or adaptive expert reasoning. As AI agents increasingly participate in educational ecosystems, dashboards designed solely for behavioural visualisation become insufficient for supporting AI-native pedagogical intelligence. This paper argues that future educational dashboards should evolve beyond behavioural visualisation toward cognition-centred infrastructures capable of representing interpretable cognitive structures for both human and AI reasoning.

\subsection{AI Tutors, Learner Modelling, and Adaptive Personalisation}

Intelligent tutoring systems have a long history of modelling learner knowledge and adapting instruction accordingly. Early cognitive tutors demonstrated how rule-based models of learner performance could support step-level guidance, feedback, and personalised learning support \cite{anderson1995cognitive,ritter2007cognitive,aleven2010rule}. Comparative work has also shown that intelligent tutoring systems can achieve meaningful learning gains relative to other forms of tutoring support \cite{vanlehn2011relative}.

A central component of these systems is learner modelling. Learner and skill models enable tutoring systems to estimate student knowledge, diagnose performance, and adapt instructional pathways \cite{desmarais2012review}. Open learner models further extend this tradition by making aspects of the learner model visible to students and teachers, supporting reflection, self-assessment, and learner agency \cite{bull2010open}. Large-scale tutoring ecosystems such as ASSISTments demonstrate how adaptive platforms can support both classroom learning and research on human learning processes \cite{heffernan2014assistments}.

AI tutoring has also expanded beyond rule-based adaptation toward conversational, affect-aware, and dialogue-based support. AutoTutor and Affective AutoTutor, for example, show how learners can interact with systems that provide conversational feedback and respond to cognitive and emotional dimensions of learning \cite{d2013autotutor}. More recently, large language models have renewed interest in AI-based feedback and tutoring, including work that connects LLM-based feedback with insights from intelligent tutoring systems and the learning sciences \cite{stamper2024enhancing}. OLiMent similarly illustrates how conversational agents can be connected with open learner modelling to help learners understand and self-assess learning goals \cite{yuan2025oliment}.

Despite these advances, AI Tutors and adaptive learning systems often remain grounded in learner performance, interaction traces, skill estimates, or conversational responses. They can personalise instruction and provide timely support, but they do not necessarily provide a broader cognition infrastructure through which learner behaviours are transformed into interpretable cognition structures for human and AI reasoning. This paper builds on the traditions of intelligent tutoring, learner modelling, and open learner modelling, but shifts the focus from adaptive personalisation alone toward cognition-centred reporting and AI-readable cognition representation.

\subsection{Human--AI Reasoning, Explainability, and Shared Interpretation}

As AI systems become more deeply embedded in educational decision-making, their value depends not only on predictive accuracy or automated support, but also on whether their reasoning can be understood, inspected, and appropriately governed. Human-centred AI emphasises the design of intelligent systems that augment rather than replace human judgment, while remaining understandable, controllable, and accountable to users \cite{shneiderman2022human}. In parallel, guidelines for human--AI interaction highlight the importance of making AI systems intelligible, controllable, and responsive to human correction during use \cite{amershi2019guidelines}.

Research on explainable AI further shows that explanations are not merely technical descriptions of model behaviour, but social and interpretive processes shaped by human goals, contexts, and expectations \cite{miller2019explanation}. In educational settings, this suggests that AI systems should not only produce outputs, but also provide interpretable accounts of how learner states, risks, or intervention needs are inferred. Without such interpretability, AI-supported educational systems risk becoming opaque decision tools rather than collaborative pedagogical systems.

Value-sensitive design provides an important foundation for this work by arguing that human values should be considered throughout the design of technical systems \cite{friedman1996value}. For AI-native education, this means that cognition infrastructures should be designed with attention to learner agency, transparency, contestability, privacy, and pedagogical responsibility. Dashboards and AI reports should not simply make learners more observable; they should support meaningful human interpretation and ethical educational action.

Suchman's work on human--machine reconfigurations also cautions against treating interaction with technical systems as a simple transfer of plans or decisions between humans and machines \cite{suchman2007human}. Instead, human--AI reasoning should be understood as situated, relational, and shaped by the contexts in which people and systems act together. This perspective is important for education because learner interpretation, teacher judgment, and AI-generated recommendations are all embedded within situated pedagogical practices.

However, existing work on human-centred, explainable, and value-sensitive AI does not fully specify the educational cognition representations needed for shared reasoning between learners, educators, AI Tutors, and AI Twins. The challenge is not only to explain AI outputs, but to create shared cognition structures through which human and AI actors can jointly reason about learner development. This paper addresses this gap by positioning the ECD as a cognition-sharing infrastructure that supports interpretable, reviewable, and AI-readable educational reasoning.

\subsection{Expert Cognition, Tacit Knowledge, and Situated Judgment}

Expert knowledge is not limited to explicit rules, procedures, or declarative content. Polanyi's account of tacit inference argues that people often know more than they can fully articulate, suggesting that expertise involves forms of judgment and recognition that are difficult to reduce to explicit statements \cite{polanyi1966logic}. This is especially important for AI-native education because the most valuable aspects of expert reasoning may lie not only in what experts know, but in how they notice patterns, interpret situations, recognise tensions, and decide what matters.

Research on expertise similarly emphasises that expert performance is shaped by experience, perception, and situated judgment. Dreyfus and Dreyfus describe expertise as progressing beyond rule-following toward intuitive and context-sensitive action \cite{dreyfus1986mind}. Klein's work on naturalistic decision-making further shows that experts often make decisions by rapidly recognising meaningful patterns in complex situations rather than by exhaustively comparing formal alternatives \cite{klein2017sources}. The broader expertise literature also highlights that expert performance depends on domain-specific knowledge, pattern recognition, deliberate practice, and the ability to act effectively under contextual constraints \cite{ericsson2018cambridge}.

Professional expertise is also reflective and interpretive. Sch{\"o}n's account of the reflective practitioner shows that professionals think in action, responding to uncertain and changing situations through reflection, reframing, and situated judgement \cite{schon2017reflective}. This perspective is highly relevant to educational practice, where expert educators do not simply apply fixed rules to learner data. Rather, they interpret learner explanations, hesitation, misunderstanding, confidence, identity development, and contextual cues in order to decide how to intervene.

These ideas are reinforced by work on situated cognition and situated learning. Brown, Collins, and Duguid argue that knowledge is inseparable from the activity, context, and culture in which it is used \cite{brown1989situated}. Lave and Wenger similarly frame learning as participation in situated social practice rather than the acquisition of abstract knowledge alone \cite{lave1991situated}. From this perspective, both learning and expert judgment are embedded in practice, interaction, and context.

The difficulty of extracting expert knowledge has also been recognised in knowledge engineering. Hoffman notes that expert knowledge is often hard to elicit because experts may not be able to fully verbalise the perceptual cues, distinctions, and reasoning processes that guide their decisions \cite{hoffman1987problem}. This challenge is directly relevant to educational AI: if expert cognition is tacit, situated, and interpretive, then AI systems cannot rely solely on explicit rules or behavioural indicators to support expert-like reasoning.

The ECD builds on this insight by treating learner behaviour as the starting point rather than the endpoint of educational interpretation. Its aim is not to fully capture or automate expert judgment, but to provide cognition structures that make aspects of learner interpretation, misunderstanding, value recognition, identity development, and cognitive tension more visible and reviewable. In this way, ECD seeks to support human--AI educational reasoning by creating representations that are closer to the forms of situated judgment used by experts than conventional behavioural dashboards alone.

The ECD proposed in this paper is informed by these perspectives on tacit knowledge and interpretive cognition. Rather than treating learning as a purely behavioural phenomenon, the framework emphasises cognition structures such as interpretation, identity cognition, value recognition, and cognitive tension as essential dimensions for AI-driven educational reasoning. In this sense, cognition-aware dashboards provide a potential infrastructure for translating tacit expert reasoning into interpretable educational intelligence within AI-native learning systems.

\section{From Learning Analytics to Cognition intelligence}

Current educational AI systems largely rely on behavioural analytics as the primary mechanism for understanding learning processes. Metrics such as correctness, completion rates, interaction frequency, and engagement duration are commonly used to evaluate learner performance and generate adaptive recommendations. While these approaches have improved scalability and automation in education, they remain limited in their ability to represent the deeper cognitive dynamics underlying learner development.

This paper argues that the future of AI-native education requires a paradigm shift from behavioural analytics toward Cognition intelligence. Rather than treating learning as a collection of observable activities, cognition-centred systems seek to interpret how learners construct meaning, negotiate uncertainty, develop identity, recognise value, and experience conceptual tension. Under this perspective, educational dashboards should not merely display behavioural statistics, but should function as infrastructures for cognition interpretation and expert reasoning.

\subsection{Why Behaviour Is Not Enough}

Behavioural data provides useful indicators of learner activity, but behaviour alone cannot fully represent cognition. Observable actions such as answering questions, completing tasks, or interacting with interfaces often reveal only the surface manifestation of underlying cognitive processes. Two learners may produce identical behavioural outcomes while possessing fundamentally different interpretations, conceptual structures, motivations, or reasoning pathways.

For example, a learner may provide an incorrect answer due to factual misunderstanding, interpretive ambiguity, conceptual conflict, value-related hesitation, or uncertainty about contextual assumptions. Traditional learning analytics systems typically register such cases as equivalent behavioural errors, even though their cognitive origins differ significantly. As a result, behaviour-centred systems frequently oversimplify learner development by reducing cognition to measurable outputs.

Human experts rarely interpret learning solely through behavioural observation. Experienced educators often infer deeper cognitive states through patterns of explanation, hesitation, conceptual framing, emotional response, and interpretive consistency. Expert pedagogical reasoning depends not only on what learners do, but also on how learners understand, interpret, and construct meaning within a learning context.

This distinction highlights a fundamental limitation in current AI educational systems. Although modern AI Tutors can process large quantities of learner interactions, they generally lack infrastructures for transforming behaviours into expert-readable cognition structures. Without cognition-level interpretation, AI systems remain constrained to reactive personalisation rather than expert-like educational reasoning.

Consequently, future AI-native education systems require mechanisms capable of translating behavioural traces into interpretable cognitive representations. This transition forms the foundation for moving beyond traditional learning analytics toward Cognition intelligence.

\subsection{Cognition Intelligence}

This paper introduces the concept of \textit{Cognition Intelligence} as a new paradigm for AI-driven educational interpretation. The term refers to a shift from analysing learner behaviour as the primary object of educational intelligence toward modelling the expert-informed cognition structures through which learner development can be interpreted.

\begin{quote}
\textbf{Definition.} Cognition Intelligence refers to the transformation of learner behaviours into interpretable cognition structures that support expert-level reasoning, adaptive intervention, and educational decision-making.
\end{quote}

Unlike traditional learning analytics, which primarily focuses on behavioural measurement and statistical visualisation, Cognition Intelligence emphasises interpretive cognition. Its objective is not merely to determine whether learners perform correctly, but to support reasoning about how learners conceptualise knowledge, negotiate meaning, construct identity, recognise value, and respond to cognitive tension.

Under this framework, learner behaviours are treated as observable signals that require interpretation rather than as direct representations of cognition itself. AI Tutors, reporting systems, and cognition-aware dashboards collectively participate in transforming behavioural traces into higher-level cognition structures. These structures are designed to make learner development more interpretable for human educators and more usable for AI systems that perform expert-like pedagogical reasoning.

Cognition Intelligence also redefines the role of educational dashboards. Rather than functioning solely as visualisation interfaces, dashboards become cognitive middleware infrastructures that mediate between learner behaviours and AI-driven expert interpretation. This enables AI systems to reason through cognition-aware representations instead of isolated behavioural statistics.

Importantly, Cognition Intelligence does not replace behavioural analytics; rather, it extends and reorganises behavioural analytics within a cognition-centred framework. Behavioural data remains necessary, but it becomes only the first layer in a larger interpretive process aimed at modelling learner development through expert-informed cognition structures.

\subsection{Cognition Structures}

Within the proposed framework, dashboards analyse not only observable behaviours but also higher-order cognition structures that emerge through interpretation. These cognition structures form the core analytical objects of the ECD.

One important dimension is \textit{identity cognition}, which refers to how learners position themselves within a domain of knowledge or practice. Identity cognition includes confidence formation, role perception, disciplinary belonging, and the development of self-recognition as a learner or practitioner. Educational growth often involves identity transformation rather than simple skill acquisition.

A second dimension is \textit{interpretation}. Learning frequently depends on how learners interpret concepts, contextualise information, and construct meaning across different situations. Two learners may memorise identical information while interpreting its significance in fundamentally different ways. Interpretive variation therefore represents an essential component of cognition-aware educational analysis.

A third dimension is \textit{cognitive tension}. Tension emerges when learners encounter conceptual conflicts, uncertainty, competing perspectives, or incompatible reasoning structures. Such tensions often indicate important moments of cognitive transition and deeper learning development. Traditional behavioural analytics rarely capture these interpretive conflicts despite their educational significance.

A fourth dimension is \textit{value recognition}. Learning is not entirely neutral or procedural; learners continuously negotiate priorities, preferences, ethical considerations, and judgments about relevance or importance. Value structures influence how learners make decisions, interpret information, and engage with educational tasks.

Finally, dashboards must model \textit{misunderstanding} not merely as incorrect performance, but as structured cognitive divergence. Misunderstandings may originate from conceptual gaps, interpretive mismatches, contextual assumptions, or unresolved tensions. Cognition-aware systems therefore seek to analyse why misconceptions emerge rather than simply detecting whether they occur.

Together, these cognition structures enable dashboards to represent learner development at a deeper interpretive level. Instead of reducing learning to behavioural outputs, cognition-centred systems provide structured representations that can support expert reasoning, adaptive intervention, and AI-native educational intelligence.

\subsection{Paradigm Shift}

The transition from learning analytics to Cognition intelligence represents a broader paradigm shift in AI education. Traditional educational systems primarily organise learner information through behavioural measurement, statistical aggregation, and visualisation interfaces designed for human instructors. In contrast, cognition-centred systems introduce interpretive infrastructures that support both human and AI-driven reasoning.

\begin{figure}[t]
    \centering
    \includegraphics[width=\linewidth]{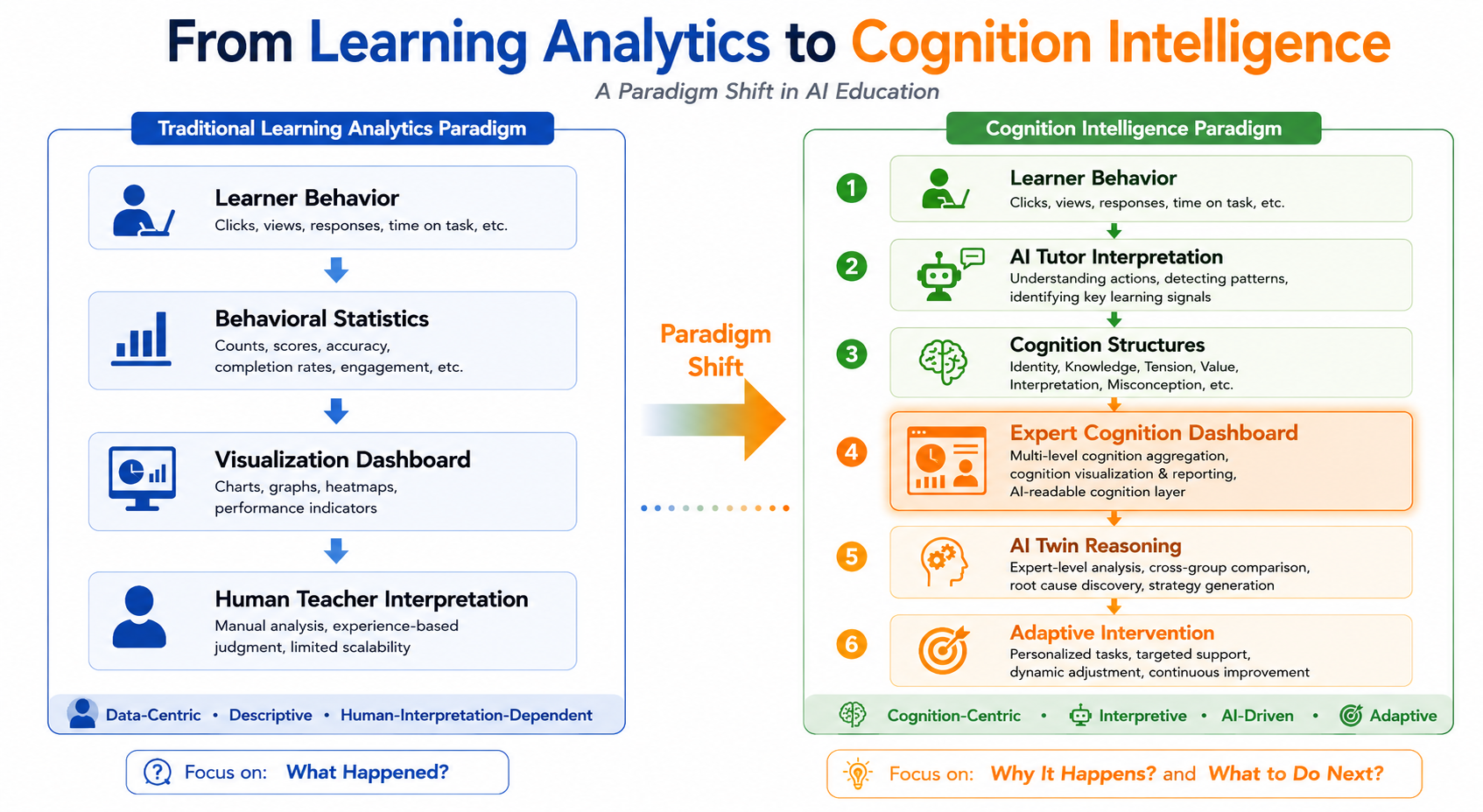}
    \caption{Paradigm shift from learning analytics to Cognition intelligence. Traditional learning analytics transforms learner behaviours into behavioural statistics and dashboard visualisations for human interpretation. In contrast, Cognition intelligence introduces AI Tutor analysis and cognition structures as an intermediate interpretive layer, enabling Expert Cognition Dashboards to support AI Twin reasoning and adaptive educational intervention. \textcopyright\ 2026 Annie Yihong Yuan. All rights reserved.}
    \label{fig:learning-analytics-to-cognition-intelligence}
\end{figure}

Figure~\ref{fig:learning-analytics-to-cognition-intelligence} illustrates this paradigm shift. Conventional learning analytics systems typically follow a pipeline in which learner behaviours are transformed into behavioural statistics and subsequently visualised through dashboards for human interpretation. While effective for monitoring activity and performance, such systems remain limited in their representation of deeper cognition structures.

The proposed framework introduces an alternative cognition-centred pipeline. In this model, learner behaviours are first interpreted through AI Tutor analysis before being transformed into cognition structures such as interpretation, identity cognition, value recognition, tension, and misunderstanding. These structures are then organised within the ECD, enabling AI Twins to perform expert-like reasoning and generate adaptive educational interventions.

Under this perspective, dashboards evolve from passive visualisation tools into active cognition infrastructures. Their role is no longer limited to displaying information, but extends toward supporting cognition interpretation, AI reasoning, and adaptive educational decision-making. This shift marks the emergence of Cognition intelligence as a foundational principle for future AI-native education systems.

\section{Expert Cognition Dashboard (ECD)}

The Expert Cognition Dashboard (ECD) is the central infrastructure proposed in this paper. Unlike conventional educational dashboards that primarily visualise behavioural statistics, ECD is designed as a cognition-centred infrastructure that transforms learner behaviours into interpretable cognitive structures for AI-driven expert reasoning and adaptive intervention.

In its software implementation, the ECD is realised as the ECD System (ECDS), a cognition-centred educational infrastructure that functions as cognitive middleware between learner behaviours and AI Twin reasoning systems. This section introduces the conceptual definition of ECD, explains its function as cognitive middleware, presents the three-level dashboard architecture, and discusses the emergence of AI-readable cognition reporting as a new paradigm for AI-native education systems.

\subsection{What Is ECD?}

Traditional educational dashboards are typically designed as visualisation interfaces that summarise learner performance through measurable indicators such as scores, engagement rates, completion statistics, and attendance records. Their primary purpose is to support monitoring and reporting for human instructors and institutions.

In contrast, the ECD redefines the role of dashboards within AI-native education ecosystems.

\begin{quote}
\textbf{Definition.} The Expert Cognition Dashboard (ECD) is a cognition-centred infrastructure that transforms learner behaviours into interpretable cognitive structures for AI-driven expert reasoning, adaptive intervention, and multi-level educational interpretation.
\end{quote}

Under this framework, dashboards are no longer limited to displaying behavioural information. Instead, they function as infrastructures for cognition construction and interpretation. The purpose of ECD is not simply to visualise what learners do, but to represent how learners think, interpret, negotiate meaning, experience tension, and develop identity within learning processes.

This distinction fundamentally changes the role of educational dashboards. In conventional systems, dashboards serve primarily as observational tools. In ECD, dashboards become active cognition infrastructures that mediate between learner behaviours and AI-driven reasoning systems. As a result, dashboards shift from passive data presentation toward cognition-aware educational intelligence.

The proposed framework also reflects a broader transition from behaviour-centred education toward cognition-centred AI systems. Instead of treating learner activity as the final analytical object, ECD positions behavioural data as the initial layer within a larger interpretive cognition pipeline. AI Tutors, cognition reports, and dashboard aggregation mechanisms collectively participate in transforming raw learning interactions into structured cognition representations that can support expert-level reasoning.

Importantly, ECD is not intended to replace human educators. Rather, it provides an infrastructure through which both humans and AI systems can access interpretable cognition structures that are otherwise difficult to observe directly through behavioural analytics alone.

\subsection{Dashboard as Cognitive Middleware}

A central argument of this paper is that educational dashboards should function as cognitive middleware rather than as visualisation endpoints. Traditional dashboards typically operate at the final stage of data pipelines. Behavioural data is collected, aggregated, and visualised for human observation. In such systems, dashboards primarily communicate statistical summaries without performing deeper cognition-level interpretation.

ECD introduces an alternative architecture in which dashboards become an intermediate cognition layer situated between learner behaviours and AI-driven expert reasoning. Within this model, dashboards do not merely present information; they actively organise, structure, and translate cognition representations that can support adaptive educational decision-making.

This cognitive middleware layer performs several critical functions. First, it transforms fragmented behavioural signals into coherent cognition structures. Learner activities such as responses, revisions, hesitations, interaction sequences, and questioning patterns are interpreted through AI Tutor analysis and reorganised into higher-level cognition representations.

Second, cognitive middleware enables multi-level aggregation. Individual cognition reports can be synthesised into class-level cognition structures, which can subsequently support cross-group reasoning at the AI Twin level. This aggregation process allows AI systems to identify collective misunderstandings, recurring tensions, interpretive patterns, and emerging learning trajectories across educational ecosystems.

Third, cognitive middleware provides interpretability for AI reasoning processes. Rather than exposing raw behavioural logs directly to AI Twins, ECD delivers cognition-aware representations that support expert-like interpretation and adaptive planning. In this sense, dashboards become perception infrastructures through which AI systems understand learners.

This shift is particularly important for AI-native education systems. Future AI educational agents will require structured cognition representations capable of supporting reasoning beyond statistical prediction. By functioning as cognitive middleware, ECD enables educational AI systems to reason through interpreted cognition structures rather than isolated behavioural data.

\subsection{Three-Level Dashboard Architecture}

The proposed ECD framework organises cognition interpretation across three interconnected layers: the Individual Cognition Dashboard, the Class Cognition Dashboard, and the AI Twin Expert Dashboard. Together, these layers form a multi-level cognition architecture that supports both local and ecosystem-level educational reasoning.

\begin{figure}[t]
    \centering
    \includegraphics[width=\linewidth]{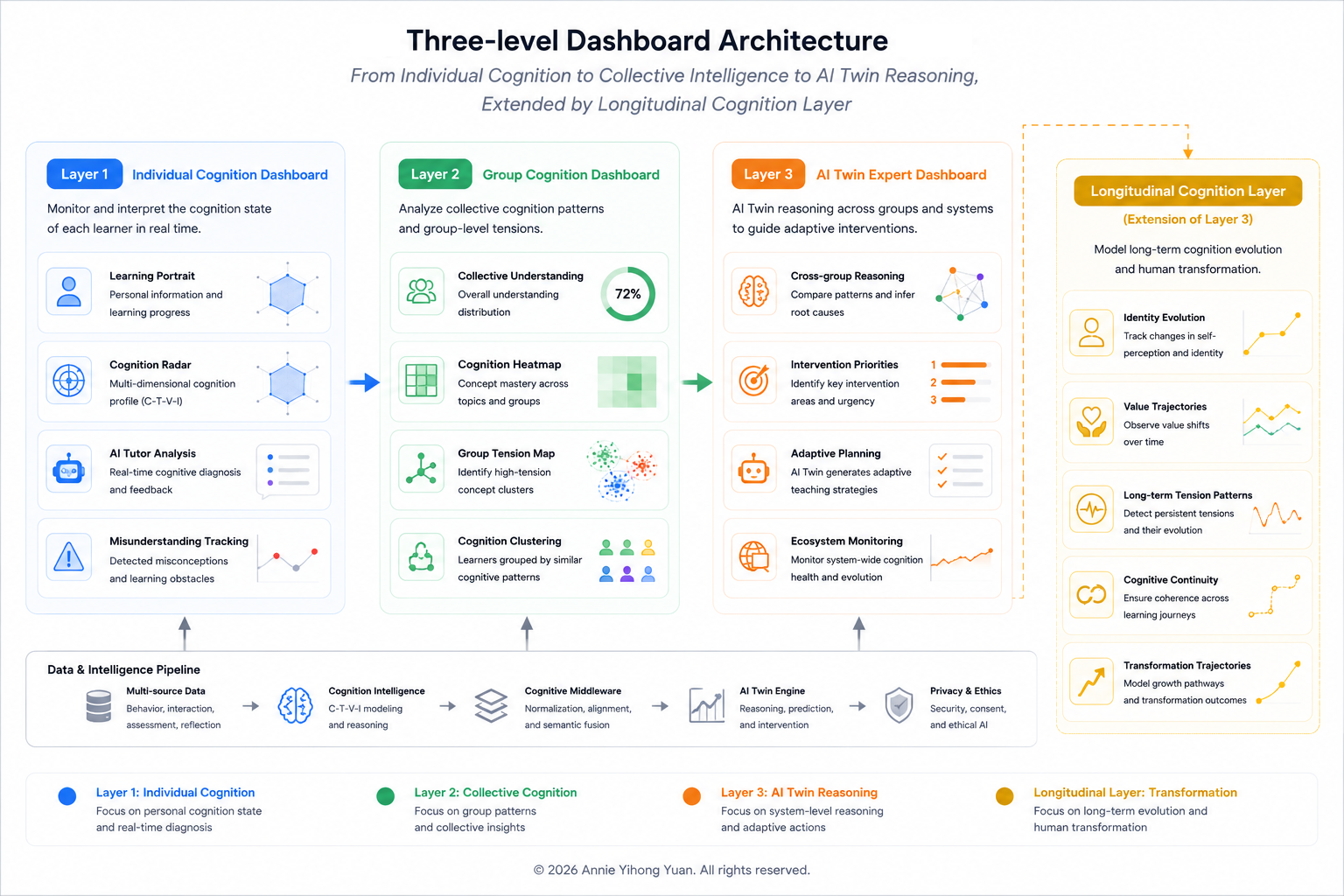}
    \caption{Three-level architecture of the Expert Cognition Dashboard. Learner behaviours are interpreted through AI Tutor analysis and transformed into cognition structures, which are organised across individual cognition dashboards, class cognition dashboards, and AI Twin expert dashboards for multi-level reasoning and adaptive intervention. \textcopyright\ 2026 Annie Yihong Yuan. All rights reserved.}
    \label{fig:ecd-three-level-architecture}
\end{figure}

Figure~\ref{fig:ecd-three-level-architecture} illustrates the overall architecture of the three-level dashboard system. The architecture shows how learner behaviours are first interpreted through AI Tutor analysis and transformed into cognition structures before being aggregated across individual, class, and AI Twin levels. This design positions the dashboard not as a final visualisation layer, but as a cognition infrastructure that enables multi-level interpretation and adaptive intervention.

\subsubsection{Individual Cognition Dashboard}

The Individual Cognition Dashboard focuses on learner-level cognition interpretation. Its primary objective is to construct a personalised cognition portrait that reflects how a learner interprets concepts, responds to tasks, develops understanding, and navigates cognitive challenges over time.

One key component is the learning portrait, which summarises learner trajectories, interaction patterns, conceptual development, and behavioural tendencies across educational activities. Unlike traditional learner profiles that emphasise performance metrics alone, the learning portrait seeks to represent evolving cognition patterns and interpretive characteristics.

Another important component is the cognition radar, which visualises multiple cognition dimensions such as conceptual understanding, interpretive consistency, identity confidence, value recognition, and cognitive tension. The cognition radar provides a structured representation of learner cognition states rather than a simple performance summary.

The Individual Dashboard also integrates AI Tutor analysis. AI Tutors continuously interpret learner behaviours and generate cognition-oriented reports that describe misconceptions, interpretive shifts, unresolved tensions, and adaptive learning needs. These reports provide the foundational cognition inputs for higher-level dashboard aggregation processes.

Together, these elements allow the Individual Dashboard to function as a cognition-aware representation of learner development rather than merely a behavioural activity tracker.

\subsubsection{Class Cognition Dashboard}

The Class Cognition Dashboard aggregates individual cognition reports into collective cognition structures that represent group-level learning dynamics. Traditional class dashboards often summarise classroom activity through averages, rankings, or participation statistics. In contrast, the proposed framework focuses on cognition aggregation rather than performance aggregation.

One major component is the identification of collective misunderstandings. By synthesising cognition reports across learners, the system can detect recurring conceptual confusion, shared interpretive errors, and common misconceptions emerging within the learning environment.

Another important component is the analysis of group tension. Cognitive tension may emerge when learners collectively struggle with conceptual ambiguity, conflicting interpretations, or unstable reasoning structures. Detecting such tensions allows AI systems and educators to identify critical moments requiring adaptive intervention.

The Class Dashboard also incorporates cognition heatmaps, which visualise the distribution and intensity of cognition patterns across groups. Rather than displaying engagement intensity alone, cognition heatmaps represent interpretive concentration, misunderstanding clusters, identity instability, and conceptual divergence within collective learning processes.

By organising learner cognition at the group level, the Class Dashboard enables broader educational reflection that extends beyond individual performance measurement.

\subsubsection{AI Twin Expert Dashboard}

The AI Twin Expert Dashboard represents the highest-level cognition layer within the proposed architecture. Its purpose is to support ecosystem-level reasoning, adaptive planning, and expert-like intervention across multiple learners, groups, and educational contexts.

At this level, the AI Twin no longer observes raw learner behaviours directly. Instead, it reasons through aggregated cognition structures generated by lower-level dashboards and AI Tutor interpretation systems.

One key function is cross-group reasoning. The AI Twin can identify cognition patterns across multiple classes or learner communities, detect recurring misunderstandings, compare interpretive trajectories, and recognise systemic educational tensions.

A second function involves determining intervention priorities. By analysing cognition distributions and tension patterns, the AI Twin can identify which conceptual areas require immediate pedagogical attention, which learners need targeted support, and which groups demonstrate unstable cognition development.

The dashboard also supports adaptive planning. Based on cognition-aware analysis, AI Twins can generate personalised tasks, modify instructional strategies, adjust learning sequences, and coordinate interventions dynamically across educational environments.

Through these functions, the AI Twin Expert Dashboard transforms dashboards from reporting interfaces into infrastructures for AI-driven expert cognition and educational orchestration. Beyond short-term cognition analysis, the AI Twin Dashboard may further evolve into a longitudinal cognition layer capable of modelling identity evolution, value trajectories, persistent tensions, and long-term cognitive transformation. This extension positions the ECD not only as a learning analytics infrastructure, but also as a potential human transformation infrastructure for AI-native educational ecosystems.

\begin{figure}[t]
    \centering
    \includegraphics[width=\linewidth]{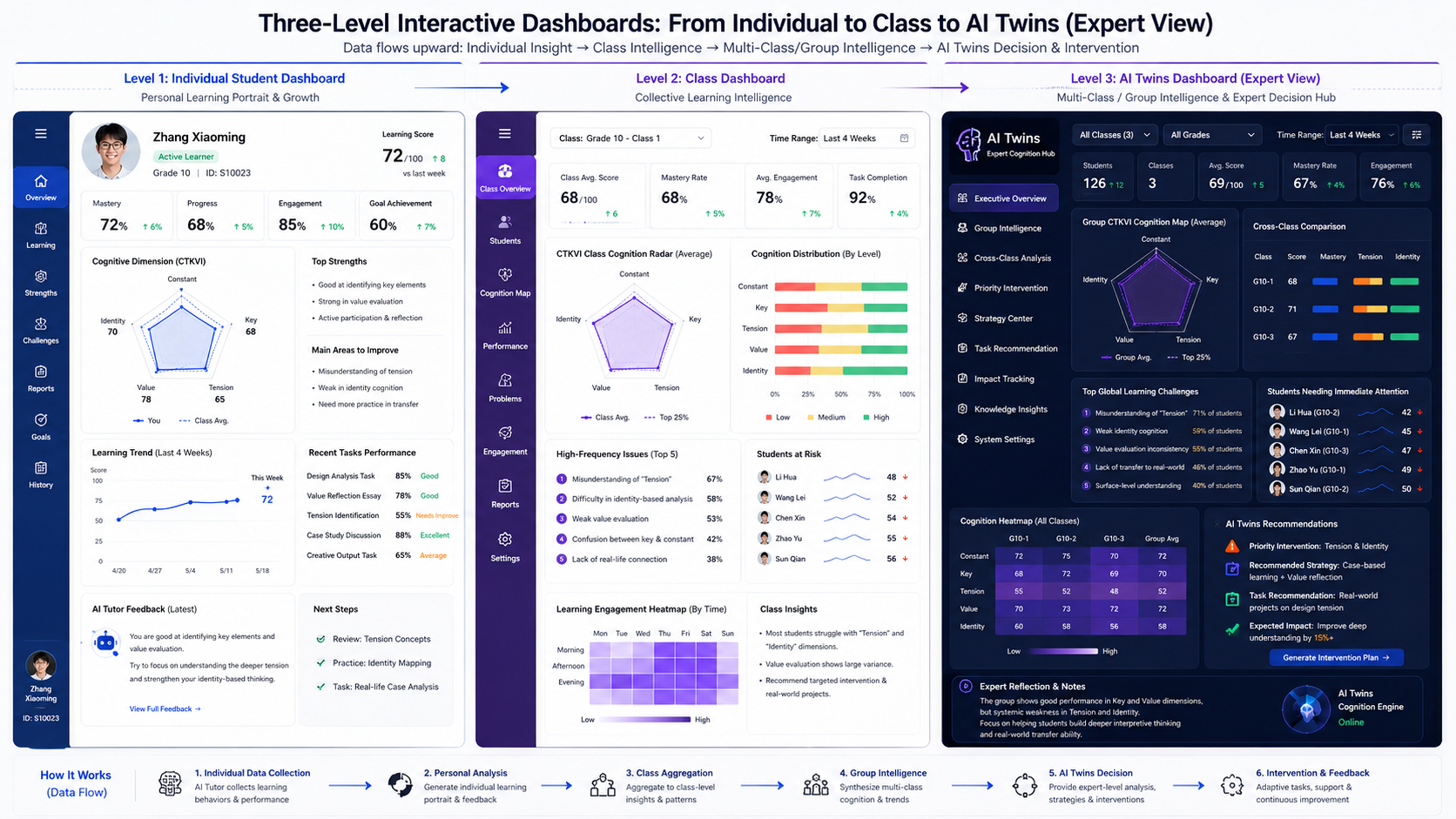}
    \caption{Prototype interface of the Expert Cognition Dashboard System. The interface demonstrates how cognition structures may be operationalised through interactive components such as learner portraits, cognition summaries, class-level cognition aggregation, and AI Twin reasoning panels. \textcopyright\ 2026 Annie Yihong Yuan. All rights reserved.}
    \label{fig:ecd-prototype-interface}
\end{figure}

Figure~\ref{fig:ecd-prototype-interface} presents a prototype interface of the ECD System using the interface design in \texttt{interactive\_dashboards.png}. The prototype illustrates how cognition-aware reporting may be operationalised through interactive dashboard components, including learner portraits, cognition summaries, class-level aggregation views, and AI Twin reasoning panels. Rather than functioning as a static visual report, the interface represents ECDS as an interactive cognition infrastructure through which learners, educators, AI Tutors, and AI Twins can access, inspect, and act upon interpreted cognition structures.

\subsection{AI-Readable Reporting}

A major implication of the proposed framework is the emergence of AI-readable cognition reporting as a new form of educational infrastructure. Traditional educational reports are primarily designed for human readability. Their objective is to communicate information visually through charts, summaries, and statistical indicators that educators can interpret manually. While effective for human monitoring, such reports are not optimised for AI-driven cognition reasoning.

In contrast, ECD introduces the concept of AI-readable cognition reports. These reports are structured not only for human interpretation, but also for machine-level reasoning and adaptive decision-making. Their purpose is to provide AI systems with interpretable cognition structures that can support expert-like educational analysis.

AI-readable reporting requires a different representational logic from conventional dashboards. Rather than emphasising raw behavioural statistics, cognition reports organise information around interpretive structures such as misunderstanding patterns, identity instability, value conflicts, conceptual tension, and learning trajectories. These structures provide meaningful cognition inputs that AI Twins can reason about directly.

This transition significantly changes the relationship between dashboards and AI systems. Dashboards no longer function merely as human-facing visualisation interfaces. Instead, they become cognition infrastructures through which AI systems perceive learners, interpret educational states, and generate adaptive interventions.

Under this perspective, AI Tutors operate as cognition interpreters, dashboards function as cognitive middleware, and AI Twins perform expert-level reasoning through aggregated cognition reports. Together, these components form the foundation of a cognition-aware educational ecosystem designed for future AI-native learning environments.

\section{AI Expert Feedback Ecology}

The ECD does not operate as an isolated visualisation system. Instead, it functions within a larger cognition-centred educational ecosystem in which AI Twins, AI Tutors, dashboards, and learners continuously interact through adaptive feedback processes. This paper defines this ecosystem as the \textit{AI Expert Feedback Ecology}.

The proposed ecology introduces a closed-loop cognition framework that connects learner behaviours, expert-informed cognition interpretation, dashboard aggregation, and AI-driven intervention into a continuous adaptive cycle. Unlike traditional educational systems that primarily rely on static evaluation and one-directional feedback, the AI Expert Feedback Ecology supports dynamic cognition-aware interaction across multiple educational layers. Within this framework, cognition is not treated as a fixed learner attribute, but as an evolving interpretive structure continuously shaped through interaction, reflection, and intervention.

\begin{figure}[t]
    \centering
    \includegraphics[width=\linewidth]{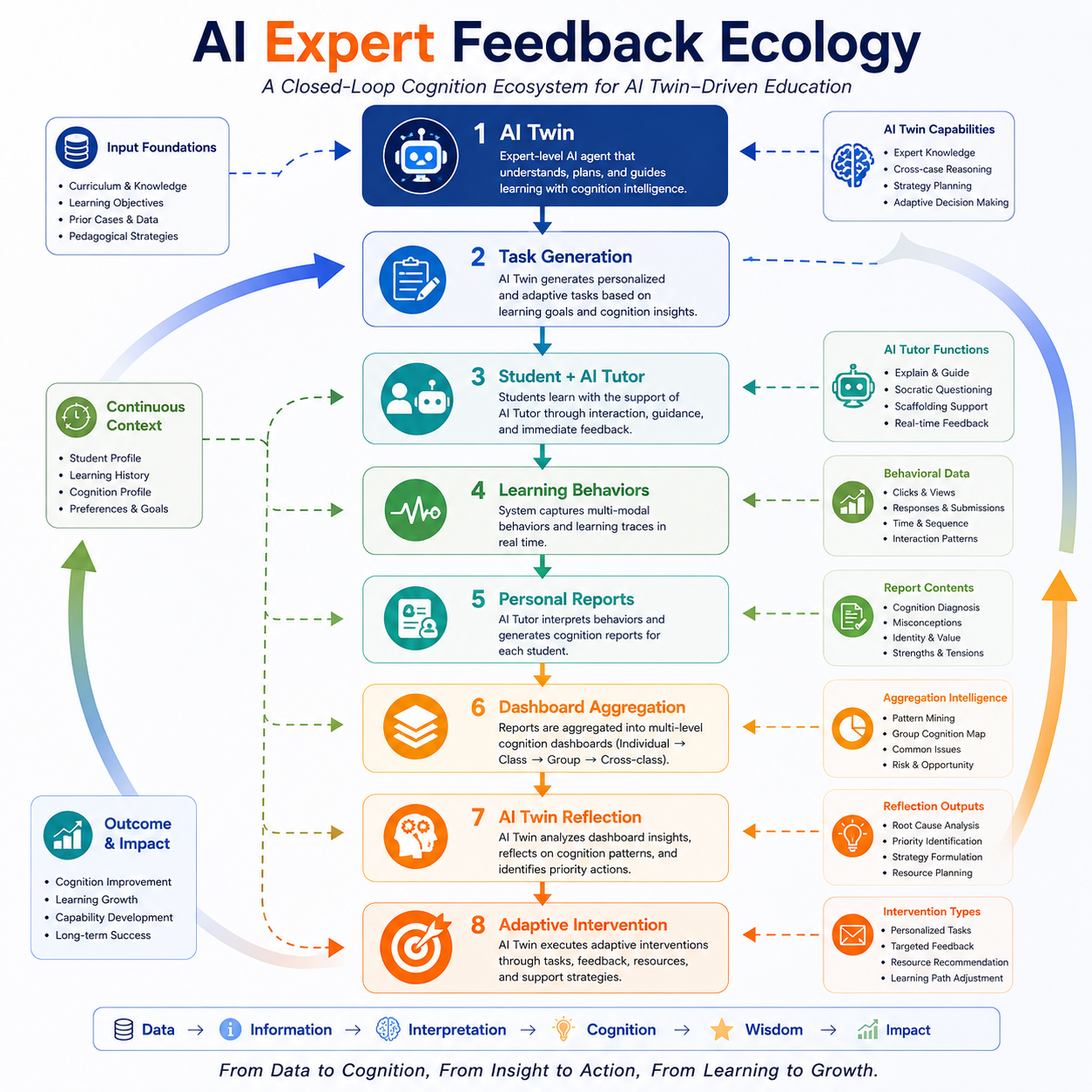}
    \caption{AI Expert Feedback Ecology. The Expert Cognition Dashboard operates within a closed-loop educational ecosystem in which AI Tutors interpret learner behaviours, dashboards aggregate cognition reports, and AI Twins reason through expert-informed cognition structures to generate adaptive interventions.\textcopyright\ 2026 Annie Yihong Yuan. All rights reserved.}
    \label{fig:ai-expert-feedback-ecology}
\end{figure}

Figure~\ref{fig:ai-expert-feedback-ecology} illustrates the overall ecology. The figure shows how learner interaction, AI Tutor interpretation, ECD aggregation, and AI Twin intervention form a continuous feedback loop for AI-native education.

\subsection{Closed-Loop Cognition Ecology}

Traditional educational feedback systems are often linear. Learners complete tasks, instructors evaluate performance, and feedback is subsequently delivered through grades, comments, or recommendations. Such systems generally separate learning activities from deeper cognition interpretation and rarely support continuous adaptive reasoning at scale.

The AI Expert Feedback Ecology introduces an alternative model based on closed-loop cognition interaction. The process begins with AI Twins generating adaptive learning tasks based on existing cognition reports and educational objectives. Learners then engage with these tasks alongside AI Tutors, which continuously interpret learner behaviours during the interaction process.

Behavioural signals generated during learning activities are transformed into personalised cognition reports through AI Tutor interpretation. These reports are subsequently aggregated within the ECD to produce group-level cognition structures and ecosystem-level cognition representations. AI Twins then reason through these aggregated cognition structures to perform reflection, identify intervention priorities, and generate adaptive educational responses. The resulting interventions produce new learning activities, thereby continuing the cognition feedback cycle.

This closed-loop architecture fundamentally changes the role of educational feedback. Feedback is no longer limited to post-task evaluation; instead, it becomes an ongoing cognition-aware process that continuously shapes educational adaptation and expert reasoning.

Importantly, the ecology also redefines how AI systems perceive learners. Rather than directly observing raw learner behaviours, AI Twins perceive learners indirectly through cognition reports generated by AI Tutors and dashboard aggregation mechanisms. In this sense, dashboards function as cognition perception infrastructures that mediate between learning activity and AI-driven educational reasoning.

The closed-loop ecology therefore represents a transition from behaviour-centred feedback systems toward cognition-centred educational ecosystems capable of supporting continuous adaptive intelligence.

\subsection{AI Tutor Interpretation}

Within the proposed framework, AI Tutors serve as the primary cognition interpretation agents responsible for transforming learner behaviours into interpretable cognition representations.

Traditional tutoring systems often focus on content delivery, question answering, or procedural assistance. Although such systems can provide immediate feedback and personalised explanations, they frequently treat learner interactions as isolated instructional events rather than as evolving cognition processes. In contrast, the proposed AI Tutor framework emphasises continuous cognition interpretation throughout the learning process.

During task interaction, AI Tutors observe a wide range of learner behaviours, including response patterns, revisions, hesitation sequences, questioning strategies, interaction timing, conceptual transitions, and problem-solving trajectories. Rather than treating these behaviours as independent metrics, AI Tutors interpret them as signals that may reflect underlying cognition structures.

This process enables the generation of learner-level cognition portraits. For example, repeated hesitation around a particular concept may indicate unresolved cognitive tension rather than simple uncertainty. Frequent revisions may suggest interpretive instability or conceptual restructuring. Similarly, certain questioning behaviours may reveal emerging identity cognition or shifts in value recognition within the learning process.

The AI Tutor therefore functions not merely as a teaching assistant, but as a cognition interpretation layer capable of constructing structured representations of learner development. These representations are not intended to fully capture learner cognition, but to make aspects of learner interpretation, misunderstanding, tension, and development more visible for human and AI reasoning.

An important aspect of this interpretation process is temporal continuity. Cognition reports are not generated from isolated interactions alone; they emerge through longitudinal observation across multiple learning episodes. This enables AI Tutors to identify evolving cognition trajectories, recurring misunderstandings, interpretive shifts, and developing learning patterns over time.

The outputs generated by AI Tutors form the foundational cognition inputs for the ECD. Without this interpretation layer, dashboards would remain limited to behavioural visualisation rather than cognition-aware educational reasoning.

\subsection{Dashboard Aggregation}

Following AI Tutor interpretation, cognition reports are aggregated within the ECD to construct higher-level cognition representations across groups and educational ecosystems.

Traditional educational aggregation methods typically rely on statistical averaging or performance comparison. For example, classrooms are commonly represented through mean scores, participation rates, ranking distributions, or engagement summaries. While useful for institutional reporting, such aggregation methods provide limited insight into collective cognition dynamics.

The proposed framework introduces cognition aggregation rather than performance aggregation. Within the dashboard infrastructure, individual cognition reports are synthesised into collective cognition structures that represent group-level interpretive patterns, conceptual tensions, and misunderstanding distributions. This aggregation process enables the identification of cognition dynamics that may not be visible at the individual learner level.

One important function of dashboard aggregation is the detection of recurring misunderstandings. When multiple learners exhibit similar interpretive confusion or conceptual instability, the dashboard can identify these patterns as emerging cognition clusters requiring targeted intervention.

Another function is the recognition of group-level cognitive tension. Tension may emerge when learners collectively struggle with ambiguity, conflicting conceptual models, or unstable interpretive frameworks. Detecting such tensions allows AI systems and educators to identify moments where adaptive pedagogical strategies may be necessary.

Dashboard aggregation also supports cognition heatmapping and trend analysis across learning populations. Rather than visualising engagement intensity alone, cognition heatmaps can represent the distribution of misunderstanding, interpretive divergence, identity instability, and conceptual conflict across educational environments.

Importantly, aggregation does not eliminate learner individuality. Instead, it creates a multi-level cognition representation system in which individual cognition patterns remain connected to broader collective dynamics. This enables educational reasoning at both local and ecosystem levels simultaneously.

Through cognition aggregation, the dashboard evolves from a reporting interface into an active cognition intelligence infrastructure capable of supporting adaptive educational reasoning at scale.

\subsection{AI Twin Reflection and Intervention}

The final layer of the AI Expert Feedback Ecology is the AI Twin reasoning and intervention process.

Within the proposed framework, AI Twins do not directly analyse raw behavioural logs or isolated learner activities. Instead, they reason through cognition structures generated by AI Tutors and dashboard aggregation systems. This allows AI Twins to perform expert-like educational reflection rather than purely statistical optimisation.

One important capability is the identification of recurring errors and misunderstandings. By analysing aggregated cognition reports, AI Twins can detect conceptual confusion across learners or groups. Rather than treating these patterns merely as incorrect responses, the system interprets them as possible indicators of deeper cognition instability or interpretive mismatch.

AI Twins also identify cognitive tension across educational environments. Tension recognition is particularly important because moments of conceptual conflict often indicate opportunities for deeper learning transformation. For example, widespread hesitation around a concept may suggest unresolved interpretive ambiguity rather than insufficient procedural knowledge. Detecting such tensions enables AI systems to generate more meaningful adaptive interventions.

Based on these cognition structures, AI Twins can perform adaptive intervention planning. Interventions may include generating new tasks, modifying instructional sequences, providing targeted explanations, restructuring conceptual pathways, or introducing alternative interpretive perspectives. Importantly, intervention decisions are grounded in cognition interpretation rather than solely in performance prediction.

AI Twins may also support differentiated educational orchestration across multiple groups or learning contexts. For example, different classrooms exhibiting distinct cognition patterns may receive different adaptive strategies even when their behavioural performance metrics appear similar. This enables more context-sensitive educational adaptation.

Within the overall ecology, AI Twin reflection represents the emergence of AI-driven expert reasoning infrastructures capable of operating through cognition-aware educational interpretation. Rather than simply automating feedback delivery, AI Twins participate in a continuous cognition-centred cycle of observation, interpretation, reflection, and intervention.

This ecology therefore establishes a foundation for future AI-native educational systems in which cognition intelligence becomes a primary mechanism for adaptive learning and expert pedagogical reasoning.

\section{Design Implications for AI-Native Cognition Infrastructures}

The transition from learning analytics to Cognition intelligence introduces broader implications not only for AI-native educational systems, but also for the future design of long-term human transformation infrastructures. The proposed ECD framework redefines dashboards from behavioural visualisation tools into cognition-centred infrastructures capable of supporting AI reasoning, adaptive intervention, longitudinal cognition modelling, and transformation-aware human--AI collaboration.

With the introduction of the Longitudinal Cognition Layer, cognition infrastructures may evolve beyond short-term educational analytics toward systems capable of modelling identity evolution, value trajectories, persistent tensions, behavioural continuity, and long-term cognitive transformation. This shift suggests that future AI systems may increasingly function not merely as learning assistants, but as continuous cognition partners participating in long-term human development processes.

\begin{figure}[t]
    \centering
    \includegraphics[width=\linewidth]{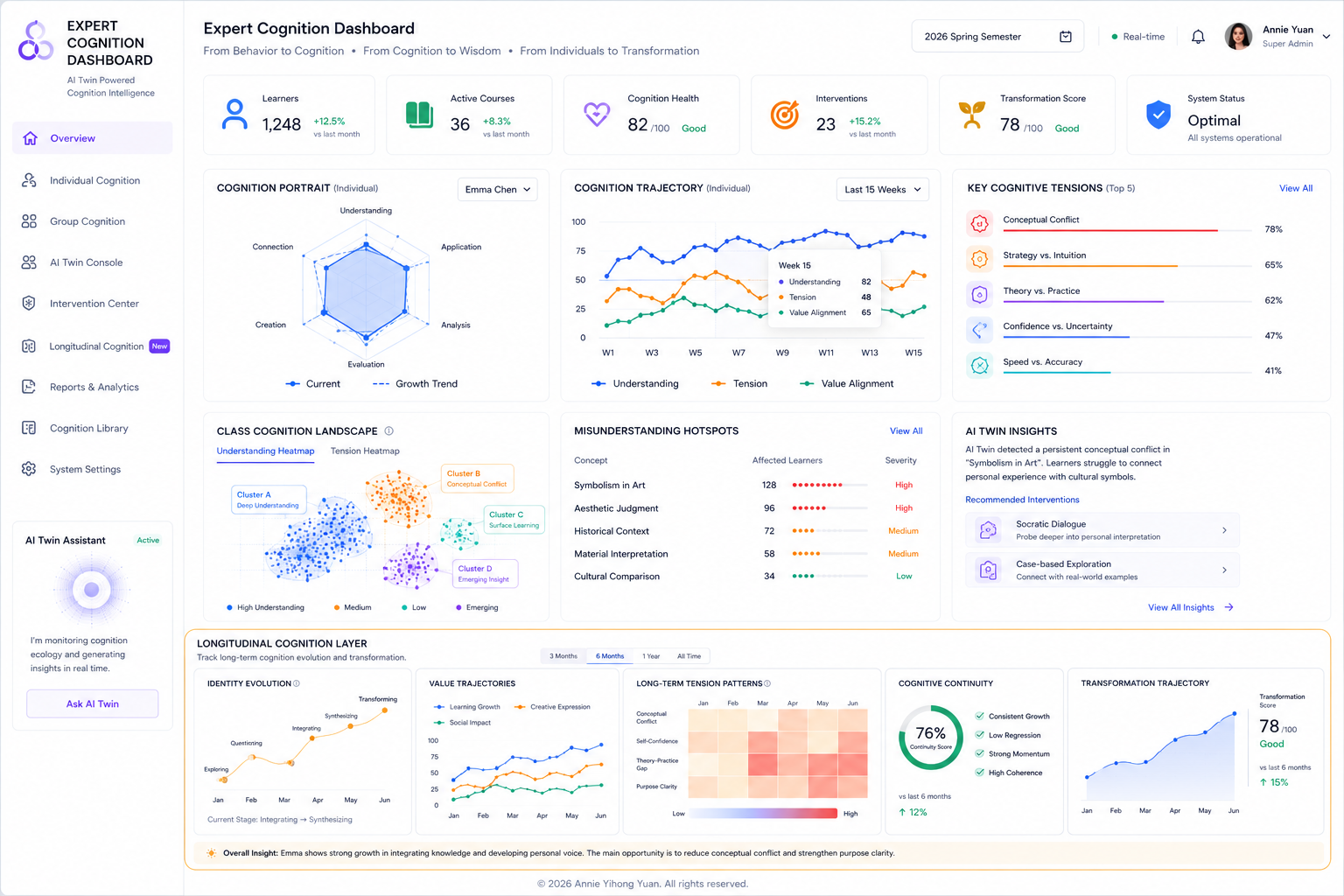}
    \caption{Expert Cognition Dashboard as a cognition-centred infrastructure. The dashboard mediates between learner behaviours, AI Tutor interpretation, cognition representation, and AI Twin reasoning, supporting adaptive intervention and multi-level educational decision-making. \textcopyright\ 2026 Annie Yihong Yuan. All rights reserved.}
    \label{fig:expert-cognition-dashboard}
\end{figure}

Figure~\ref{fig:expert-cognition-dashboard} illustrates the ECD as a cognition infrastructure that mediates between learner behaviours, AI Tutor interpretation, dashboard-level cognition representation, and AI Twin reasoning. The figure highlights the broader design shift proposed in this paper: dashboards should no longer be understood only as human-facing visualisations, but as cognition-aware infrastructures that support both human interpretation and AI-native educational reasoning.

This section discusses several major implications emerging from the proposed framework and outlines how cognition-centred infrastructures may shape the next generation of AI-native education and human transformation systems.

\subsection{From Visualisation to Cognition Infrastructure}

One of the most significant implications of this work is the transformation of dashboards from visualisation tools into cognition infrastructures.

Traditional educational dashboards are primarily designed to display behavioural statistics for human observation. Their role is largely informational: they summarise learner activity, visualise performance trends, and support institutional monitoring. Under this paradigm, dashboards function as endpoints within data pipelines.

The proposed framework fundamentally repositions dashboards within educational systems. In the ECD model, dashboards become active cognition infrastructures responsible for organising, structuring, and mediating cognition representations across educational ecosystems.

This shift changes both the technical and conceptual role of dashboards. Rather than merely presenting data, cognition infrastructures support interpretation, reasoning, adaptive intervention, and multi-level cognition aggregation. Dashboards become environments through which AI systems perceive learners, recognise tensions, identify misunderstandings, and generate educational responses.

This transition also suggests that future educational systems may increasingly depend on cognition architecture design rather than visualisation design alone. The central challenge is no longer how to display behavioural information more effectively, but how to construct interpretable cognition representations that can support both human and AI reasoning.

As AI systems become more deeply integrated into education, cognition infrastructures may become as foundational to educational AI as databases and interfaces are to contemporary software systems.

\subsection{AI-Readable Educational Systems}

A second major implication concerns the emergence of AI-readable educational systems.

Most existing educational technologies are primarily designed for human readability. Learning reports, dashboards, and analytics platforms focus on presenting information visually in ways that teachers and administrators can manually interpret. Although AI may participate in backend prediction or recommendation processes, the educational interface itself remains fundamentally human-centred.

The proposed framework introduces an alternative paradigm in which educational reporting systems are designed simultaneously for human interpretation and machine reasoning. In this model, cognition reports become structured representations that AI systems can directly interpret, compare, and reason through.

This distinction is important because AI systems require different forms of informational organisation than humans. Raw behavioural logs, isolated metrics, and fragmented activity traces often provide insufficient structure for expert-level educational reasoning. AI-readable cognition reports therefore organise educational information around higher-order cognition structures such as misunderstanding, interpretive divergence, identity instability, conceptual tension, and value recognition.

The emergence of AI-readable reporting may significantly influence the future architecture of educational systems. Future platforms may increasingly generate dual-layer educational outputs: human-readable visualisations for educators and AI-readable cognition representations for intelligent educational agents.

Under this perspective, educational dashboards become shared cognition interfaces between humans and AI systems. Such infrastructures may enable new forms of collaborative pedagogical intelligence in which human educators and AI agents reason through partially shared cognition representations.

\subsection{Cognition-Aware Learning Analytics}

The proposed framework also suggests a broader transformation of learning analytics itself.

Contemporary learning analytics largely focuses on behavioural prediction, engagement measurement, and performance optimisation. While these approaches have improved educational scalability and monitoring capabilities, they frequently overlook interpretive cognition and tacit educational reasoning.

Cognition-aware learning analytics extends the scope of analytics beyond observable activity toward deeper cognition interpretation. Instead of treating learning behaviours as the final analytical object, cognition-aware systems analyse how learners construct meaning, negotiate uncertainty, develop identity, and experience conceptual tension throughout learning processes.

This shift introduces several important methodological implications. First, educational analytics may increasingly require interpretive rather than purely statistical models. Understanding cognition structures such as tension, misunderstanding, or identity development often demands contextual interpretation that cannot be fully reduced to numerical indicators alone.

Second, future analytics systems may need to integrate longitudinal cognition trajectories rather than isolated behavioural events. Learner cognition evolves over time through repeated interaction, reflection, revision, and adaptation. Cognition-aware systems therefore require mechanisms capable of modelling dynamic interpretive development across extended educational experiences.

Third, cognition-aware analytics may enable more meaningful adaptive intervention. By identifying the cognitive origins of misunderstanding or instability, AI systems can generate interventions that address deeper conceptual structures rather than simply optimising short-term performance metrics.

This transition suggests that the future of learning analytics may increasingly depend on cognition modelling rather than behavioural monitoring alone.

\subsection{Human--AI Co-Teaching}

The ECD framework also has important implications for human--AI co-teaching systems.

Most current AI-assisted educational systems position AI primarily as a support tool for automation, recommendation, or content generation. Human educators remain responsible for deeper pedagogical interpretation, contextual reasoning, and adaptive decision-making.

The proposed framework introduces the possibility of cognition-centred collaboration between humans and AI systems. Because dashboards organise cognition structures into interpretable representations, both human educators and AI agents can participate in reasoning processes grounded in shared cognition reports.

Under this model, AI Tutors continuously interpret learner behaviours, dashboards aggregate cognition structures, and AI Twins generate adaptive reasoning and intervention proposals. Human educators can then review, modify, contextualise, or refine these interventions based on broader pedagogical judgment and ethical considerations.

Importantly, this framework does not seek to replace human teachers. Instead, it repositions educators within a cognition-augmented ecosystem where AI systems assist with large-scale cognition interpretation while humans contribute contextual awareness, emotional understanding, ethical reflection, and pedagogical expertise.

This form of human--AI co-teaching may become increasingly important as educational environments grow more personalised, large-scale, and continuously adaptive. Shared cognition infrastructures could enable collaborative reasoning processes that neither humans nor AI systems could achieve independently.

\subsection{Future AI-Native Education}

The broader implication of this work is the emergence of cognition-centred AI-native education systems.

Current educational AI systems remain heavily influenced by industrial-era educational logic, where learning is measured primarily through observable performance, standardised evaluation, and behavioural optimisation. Even advanced AI Tutors often operate within this behaviour-centred paradigm.

The proposed framework suggests an alternative future in which Cognition intelligence becomes the foundational infrastructure of AI education. In such systems, learning is interpreted not only through performance outcomes, but through evolving cognition structures including interpretation, identity formation, value negotiation, misunderstanding, and conceptual tension.

Future AI-native educational ecosystems may therefore depend less on isolated intelligent models and more on integrated cognition architectures capable of supporting continuous interpretation and adaptive reasoning across multiple educational layers.

Within this vision, AI Tutors function as cognition interpreters, dashboards operate as cognitive middleware, and AI Twins perform expert-level reasoning through aggregated cognition reports. Education becomes an evolving cognition ecology rather than a sequence of isolated instructional events.

This shift may also influence how educational success itself is defined. Rather than optimising only for correctness, efficiency, or completion, future systems may increasingly focus on interpretive development, conceptual transformation, adaptive reasoning, and cognition growth.

Ultimately, the transition from learning analytics to Cognition intelligence represents not merely a technological upgrade, but a broader reconceptualisation of how AI systems understand learning, support expertise, and participate in educational practice.

\subsection{Human State Computation Infrastructure}

The future of cognition infrastructures may extend beyond educational analytics toward long-term human transformation systems capable of modelling identity evolution, value trajectories, and longitudinal cognitive development. Existing information systems primarily manage behavioural records and static user attributes, but lack mechanisms for representing longitudinal human transformation and cognitive state evolution.

We propose a Cognitive Data Infrastructure that transforms human expressions into computable state representations through cross-domain semantic mapping and transformation trajectory modelling. Under this view, educational cognition dashboards may become an early form of a broader human state computation infrastructure: systems that model how people change across learning, work, identity formation, decision-making, and personal development over time.

\begin{figure}[t]
    \centering
    \includegraphics[width=\linewidth]{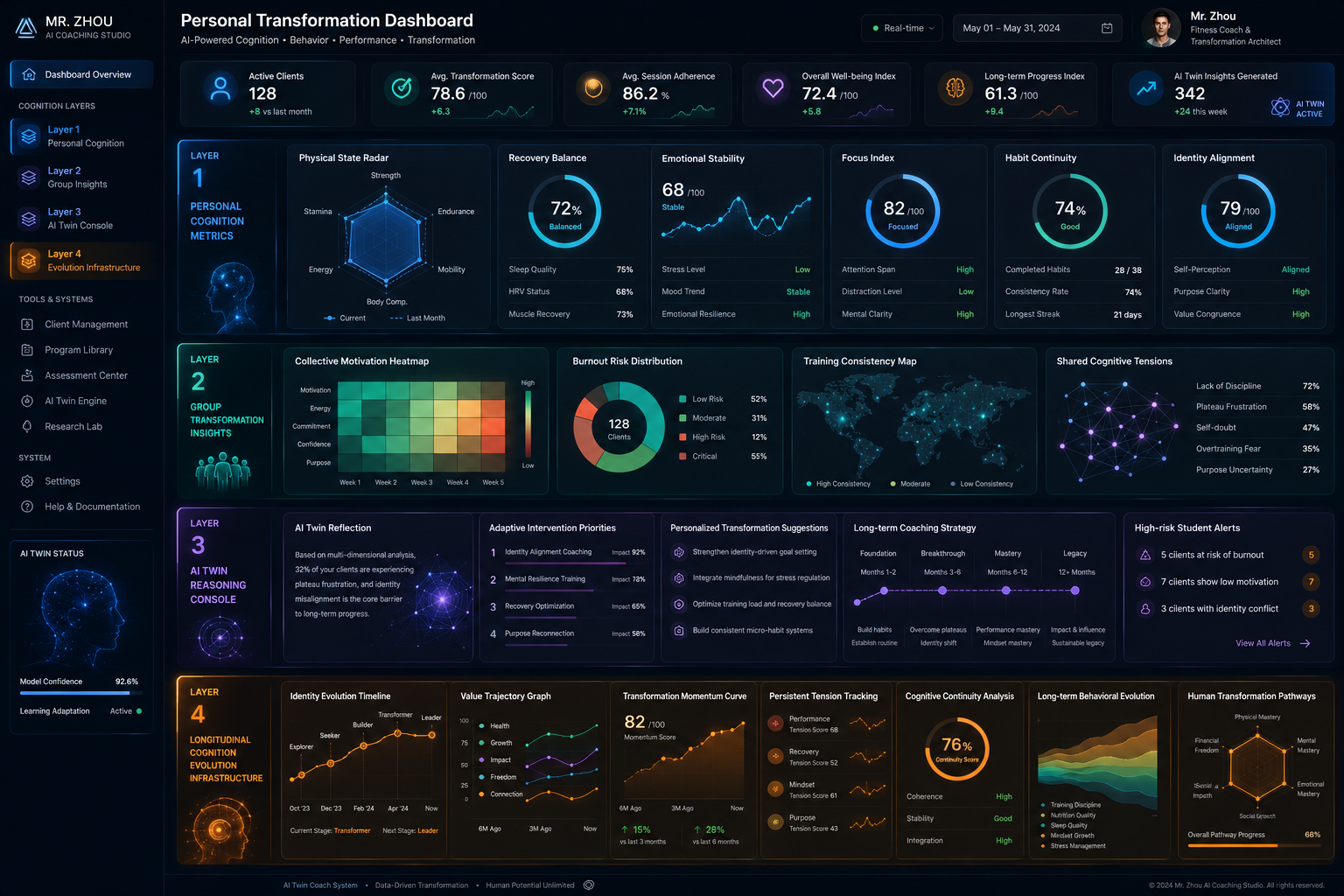}
    \caption{Personal transformation dashboard as an extension of cognition-centred infrastructure. The dashboard expands cognition modelling beyond educational analytics toward long-term representations of identity evolution, value trajectories, persistent tensions, behavioural continuity, and human transformation.\textcopyright\ 2026 Annie Yihong Yuan. All rights reserved.}
    \label{fig:personal-transformation-dashboard}
\end{figure}

Figure~\ref{fig:personal-transformation-dashboard} illustrates this broader extension from educational cognition dashboards toward personal transformation dashboards. While the ECD focuses on cognition-aware learning and educational intervention, the personal transformation dashboard expands the same logic into longer-term modelling of identity, values, tensions, goals, behavioural continuity, and developmental trajectories.

This extension raises a new design agenda for AI-native systems. Rather than only responding to immediate user actions, future AI systems may maintain longitudinal cognition models that help interpret how a person develops across contexts. Such systems could support reflection, personalised guidance, adaptive learning, professional growth, and long-term transformation-aware human--AI collaboration.

However, this broader vision also introduces significant ethical and design challenges. Longitudinal cognition infrastructures would require careful attention to transparency, consent, data ownership, interpretive accountability, and user agency. If AI systems are to participate in modelling human transformation, they must do so in ways that remain explainable, contestable, and aligned with human autonomy.

The ECD therefore represents both a concrete educational framework and a conceptual step toward broader cognition infrastructures for AI-native human development. Its central contribution is not simply a new dashboard design, but a shift in how dashboards, AI Tutors, and AI Twins may collectively support cognition-aware reasoning over time.

\section{Discussion}

The ECD framework proposed in this paper repositions educational dashboards as cognition infrastructures rather than visualisation endpoints. Across the previous sections, the paper has argued that AI-native educational systems require more than behavioural analytics, performance prediction, and engagement monitoring. They require infrastructures capable of transforming learner behaviours into interpretable cognition structures that can support expert-like reasoning, adaptive intervention, and human--AI educational collaboration.

This discussion synthesises the core theoretical and design implications of the framework. It clarifies the role of ECD as cognitive middleware, considers its implications for human--AI educational reasoning, and outlines limitations and future research directions. In keeping with the tighter framing of the paper, the discussion focuses primarily on AI-native education, while briefly noting possible extensions beyond education as future work.

\subsection{Positioning ECD as Cognitive Middleware}

A central contribution of this paper is the reframing of dashboards as cognitive middleware. In conventional learning analytics systems, dashboards are usually positioned at the end of the data pipeline. Learner behaviours are collected, processed, aggregated, and then visualised for human interpretation. In this model, dashboards primarily support visibility: they make learner activity, performance, or engagement easier to observe.

The ECD framework proposes a different role for dashboards. Rather than functioning only as visual summaries of behavioural data, dashboards become intermediate cognition infrastructures situated between learner behaviour, AI Tutor interpretation, and AI Twin reasoning. This middleware position is significant because it changes the function of dashboards from representation to mediation. The dashboard does not merely display what has happened; it organises cognition structures that shape how AI systems and human educators interpret learner development.

This reframing is particularly important for AI-native education systems. If AI Tutors and AI Twins reason directly from raw behavioural traces, they risk reducing learning to observable actions and short-term performance signals. A cognition middleware layer provides a structured interpretive buffer between behaviour and AI reasoning. It enables learner actions to be transformed into higher-order cognition representations, such as misunderstanding patterns, interpretive tensions, identity development, value recognition, and longitudinal learning trajectories.

Under this perspective, the dashboard becomes a cognition interface shared by multiple actors. Learners may use it to reflect on their own development; educators may use it to interpret learner needs and plan interventions; AI Tutors may use it to generate cognition reports; and AI Twins may use it to reason across individuals, classes, and broader educational contexts. The ECD therefore extends the dashboard from a human-facing visualisation tool into a multi-agent cognition infrastructure for education.

This also clarifies the relationship between learning analytics and Cognition intelligence. The proposed framework does not reject behavioural analytics. Behavioural traces remain necessary because they provide the observable signals from which cognition interpretations may begin. However, in the ECD model, behavioural data is not treated as the final object of analysis. Instead, it becomes the input layer for a broader cognition-centred architecture. Cognition intelligence therefore extends learning analytics by reorganising behavioural data within an interpretive framework designed to support expert-level educational reasoning.

\subsection{Implications for Human--AI Educational Reasoning}

The ECD framework has implications for how human educators and AI systems may collaborate in future educational environments. Current AI-supported education often separates human and machine roles in relatively simple ways. AI systems automate feedback, recommend resources, generate explanations, or predict learner risk, while human educators remain responsible for deeper interpretation, ethical judgment, and contextual decision-making. Although this division is useful, it can limit richer forms of human--AI collaboration.

By introducing cognition-aware dashboards, the proposed framework creates a shared representational space through which humans and AI systems can reason about learner development. Instead of AI systems producing isolated recommendations or opaque predictions, ECD structures learner cognition into interpretable forms that can be inspected, questioned, revised, and acted upon. This opens the possibility of collaborative reasoning in which human educators and AI systems engage with shared cognition reports.

For example, an AI Tutor may identify a learner's repeated misunderstanding not simply as incorrect performance, but as a pattern of conceptual tension or interpretive instability. The ECD can then represent this pattern at the individual level, aggregate similar patterns at the class level, and support AI Twin reasoning about possible interventions. A human educator may review these interpretations, bring in contextual knowledge about the course, cohort, or learner, and decide whether the proposed intervention is appropriate. In this way, ECD supports a division of labour in which AI assists with large-scale cognition interpretation while humans retain responsibility for contextual, ethical, and pedagogical judgment.

This form of cognition sharing may become especially important as educational systems become more personalised and continuous. In large-scale learning environments, individual educators may struggle to track every learner's evolving conceptual state, identity development, and long-term learning tensions. AI systems can assist by maintaining cognition reports and identifying emerging patterns. However, these reports must remain interpretable and contestable. The value of ECD lies not only in making cognition available to AI systems, but also in making AI-generated cognition interpretations visible to humans.

The framework also suggests a broader design principle for AI-native education: adaptive systems should not only optimise instruction, but also support interpretive accountability. If an AI system recommends an intervention, educators and learners should be able to understand the cognition structures on which that recommendation is based. This requires dashboards that do more than display scores or predictions. They must represent the reasoning context behind adaptive decisions.

In this sense, ECD contributes to human-centred AI education by proposing an infrastructure for shared cognition rather than simple automation. It positions AI not as a replacement for expert educators, but as a participant in a cognition ecology where learners, educators, AI Tutors, dashboards, and AI Twins interact through structured cognition representations.

\subsection{Limitations and Future Work}

This paper presents the ECD as a conceptual and design framework. As such, its primary contribution is theoretical and architectural rather than empirical. The framework identifies a gap in existing learning analytics and AI education systems, proposes Cognition intelligence as an alternative paradigm, and outlines how dashboards may function as cognitive middleware within AI-native educational ecosystems. However, the framework has not yet been validated through a full empirical deployment. Future work should therefore implement and evaluate ECD in authentic educational settings.

One important limitation concerns the challenge of modelling cognition accurately. Cognition structures such as identity cognition, value recognition, conceptual tension, and interpretive divergence are complex, situated, and context-dependent. They cannot be reliably inferred from behavioural data alone without careful theoretical grounding, domain-specific interpretation, and empirical validation. Future research should investigate how these cognition structures can be operationalised, what data sources are appropriate for modelling them, and how their validity can be assessed.

A second limitation concerns the risk of over-interpretation. AI-generated cognition reports may appear precise or authoritative even when they are uncertain, incomplete, or biased. For example, an AI Tutor may infer that a learner is experiencing identity uncertainty or conceptual tension based on limited interaction data. If such interpretations are presented without uncertainty markers or opportunities for human correction, they may misrepresent learners and lead to inappropriate interventions. Future ECD systems should therefore include mechanisms for uncertainty representation, human review, learner contestability, and transparent explanation of how cognition interpretations are generated.

A third limitation relates to the generalisability of the proposed cognition dimensions. This paper discusses structures such as interpretation, identity cognition, value recognition, cognitive tension, and misunderstanding. These dimensions are intended as initial conceptual categories rather than a fixed taxonomy. Different domains, disciplines, and learning contexts may require different cognition structures. For example, professional education, creative practice, engineering design, scientific reasoning, and humanities learning may each involve distinct forms of expert cognition. Future work should examine how the ECD framework can be adapted across domains and how domain-specific expert cognition can be represented within dashboard systems.

A fourth limitation concerns ethics and governance. The proposed framework becomes especially sensitive when extended toward longitudinal cognition modelling. Modelling identity development, value trajectories, learning tensions, and long-term cognitive transformation raises significant concerns around privacy, consent, data ownership, interpretive accountability, and learner agency. Future systems must ensure that learners understand what is being modelled, how interpretations are produced, who can access them, and how they can challenge or revise those interpretations.

Future work should proceed in several directions. First, prototype ECD systems should be developed and tested in real educational contexts to examine how learners, educators, and AI systems interact with cognition-aware dashboards. Second, empirical studies should evaluate whether cognition-centred representations improve pedagogical decision-making compared with conventional learning analytics dashboards. Third, methodological work is needed to define valid and reliable approaches for modelling cognition structures from learner interactions, reflections, assessments, and AI Tutor dialogues. Fourth, design research should explore how cognition dashboards can balance AI-readability with human interpretability, ensuring that reports are useful for machine reasoning while remaining understandable and contestable for human users.

Beyond education, the ECD framework may also inform the design of broader cognition infrastructures for organisations and long-term human--AI collaboration. Future work could examine how cognition dashboards might support organisational memory, group cognition, AI Twin coordination, behavioural analytics, and emerging cognition infrastructure roles in AI-native workplaces. However, these extensions require separate theoretical development and empirical investigation beyond the scope of the present paper. The immediate contribution of this paper is therefore to provide a conceptual foundation for investigating how dashboards may evolve from behavioural visualisation tools into cognition-centred infrastructures for AI-native education.

\section{Conclusion}

This paper has introduced the Expert Cognition Dashboard (ECD) as a cognition-centred infrastructure for AI-native education systems. In contrast to conventional learning analytics dashboards, which primarily visualise behavioural indicators such as performance, engagement, and completion, ECD reframes dashboards as cognitive middleware that transforms learner behaviours into interpretable cognition structures for AI-driven expert reasoning.

The central argument of the paper is that future educational AI systems require more than behavioural analytics. While behavioural data remains valuable, it is insufficient for representing the deeper cognitive processes that shape learning, including interpretation, misunderstanding, identity development, value recognition, and cognitive tension. To address this limitation, the paper proposed the concept of Cognition intelligence: the transformation of learner behaviours into structured cognitive representations that can support adaptive intervention, expert-like reasoning, and human--AI educational collaboration.

The paper presented a multi-level architecture for ECD, consisting of individual cognition dashboards, class cognition dashboards, and AI Twin expert dashboards. Together, these layers enable cognition interpretation across learner, group, and educational ecosystem levels. Within this architecture, AI Tutors act as cognition interpreters, dashboards function as cognitive middleware, and AI Twins perform higher-level reasoning through aggregated cognition reports. This architecture positions dashboards not as passive reporting interfaces, but as active infrastructures through which AI systems and human educators can perceive, interpret, and respond to learner development.

The paper also discussed the design implications of cognition-centred educational infrastructures, including AI-readable reporting, cognition-aware learning analytics, human--AI educational reasoning, and longitudinal cognition modelling. These implications suggest that the future of AI-native education may depend not only on more powerful models but also on better infrastructures for representing, sharing, and contesting interpretations of learner cognition.

The contribution of this paper is conceptual and architectural. It does not claim that cognition can be fully captured or automated through dashboards, nor that AI systems should replace human educators. Rather, it argues that AI-native education requires new infrastructures that make cognition more visible, interpretable, contestable, and useful for both human and machine reasoning. By shifting attention from learning analytics to Cognition intelligence, the ECD offers a foundation for designing future educational systems that are not only adaptive and data-driven but also cognition-aware, human-centred, and pedagogically grounded.

\subsection*{Copyright and Trade Mark Notice}

\noindent
\textcopyright\ 2026 Annie Yihong Yuan. All rights reserved. All figures, dashboard mock-ups, interface examples, and visual materials in this paper are original works of the author unless otherwise stated.

\noindent
Expert Cognition Dashboard\texttrademark\ is a trademark claimed by Annie Yihong Yuan. 


\clearpage
\bibliographystyle{unsrtnat}
\bibliography{references}  






\end{document}